\documentclass[12pt,a4paper,showkeys]{revtex4}
\usepackage{graphicx}
\usepackage{epsfig}
\usepackage{bm}
\usepackage{amsmath}
\usepackage{amssymb}
\usepackage{graphics}
\usepackage{epstopdf}
\usepackage[linkcolor=red,citecolor=green,urlcolor=blue]{hyperref}

\begin{document}

\title{ \textbf{Lambda production in the DIS target fragmentation region}}
\author{Federico A. Ceccopieri}
\email{federico.alberto.ceccopieri@cern.ch}
\affiliation{IFPA, Universit\'e de Li\`ege,  B4000, Li\`ege, Belgium}

\begin{abstract}
\noindent
By using a recently obtained set of Lambda fracture functions, we present predictions 
for Lambda production in the target fragmentation region of Semi-Inclusive  
Deep Inelastic Scattering in CLAS@12 GeV kinematics, supplemented with a conservative error estimates.
We discuss a number of observables sensitive to the assumptions of the underlying theory  
and many of the assumptions of the proposed phenomenological model.
\end{abstract}

\maketitle

\section{Introduction}
\noindent

Hadron production in Semi-Inclusive Deep Inelastic Scattering (SIDIS) is
usually described in terms of universal parton distributions and fragmentation functions. 
Thanks to the factorisation theorem, hadronic cross sections are obtained by
convoluting short-distance partonic cross sections, calculable in pertubation theory, with such distributions.
To lowest order in the strong coupling, this mechanism is expected to describe hadron production in the so called current fragmentation region, \textsl{i.e.} the phase space region in which the struck parton hadronises.
In order to obtain a global description of the  particle production spectrum, and in particular of hadron production in the target fragmentation region, the introduction of new non-perturbative distributions is mandatory.
This issue was early realised in Ref.~\cite{trentadue_veneziano} where the concept of fracture functions was introduced.
The latter parametrise the hadronisation into the final-state hadron of the coloured spectator system which results from the removal of the scattered parton from the initial-state hadron. For this reason, their flavour and energy dependecies are expected to be significantly different from fragmentation functions which parametrise the fragmentation of a single parton into the observed hadron. Fracture functions, by construction, simultaneously encode information both on the
parton partecipating the hard scattering and on the fragmentation of the spectator system into the observed hadron.
Therefore they constitute the connection between forward particle production at small transverse momentum (\textsl{i.e.} target fragments)  and high momentum transfer processes (\textsl{i.e.} DIS).  
Although intrinsically of non-perturbative nature, the scale dependence of such distributions 
can be predicted by perturbative QCD~\cite{trentadue_veneziano}.
Fracture functions obey, in fact, DGLAP~\cite{DGLAP} inhomogeneous evolution equations
which result from the structure of collinear singularities 
in the target fragmentation region~\cite{trentadue_veneziano,graudenz}.
Moreover, a dedicated factorisation theorem~\cite{factorization_soft,factorization_coll} 
guarantees that fracture functions are universal distributions, at least in the context of SIDIS. 
Among baryons, Lambda hyperons are predominantly produced in the SIDIS target 
fragmentation region and show a significant leading particle effect, \textsl{i.e.} they carry a significant 
fraction of the incoming proton momentum. For such reasons they have been used as a case study and 
a first attempt to determine Lambda fracture functions have been recently presented in Ref.~\cite{ceccopieri_mancusi} 
by performing a simultaneous QCD fit to a variety of Semi-Inclusive Lambda production 
data collected in lepton-nucleon scattering.
In the present paper, by using this model, we present predictions for Lambda observables
in the target fragmentation region of neutral current (NC) Deep Inelastic Scattering (DIS)
focusing on CLAS@12 GeV kinematics. 
The paper is organised as follows. In Sec.~\ref{iDIS} and Sec.~\ref{SIDIS_Lambda} 
we first briefly review the Inclusive and Semi-Inclusive DIS cross sections in lepton-nucleon scattering.
In Sec.~\ref{Model_details} we review some details of the modelisation of Lambda Fracture Functions.
In Sec.~\ref{pheno} we present and discuss a number of observables sensitive to the assumptions adopted in the model
which can be used to further constrain it. Finally in Sec.~\ref{Conclusions} we summarise our results. 

\section{Inclusive DIS}
\label{iDIS}
\noindent
The deep inelastic scattering cross section of a lepton $l$ off a proton $p$ 
\begin{equation}
\label{DISprocess}
l(k)+p(P)\rightarrow l(k')+X\,,
\end{equation}
with four-momenta $k$ and $P$, respectively, is usually described in terms of the invariants:
\begin{equation}
\label{variables0}
x_B=\frac{Q^2}{2 P\cdot q} , \;\;\; y=\frac{P\cdot q}{P\cdot k}=\frac{Q^2}{(s-m_p^2) x} ,\;\;\; Q^2=-q^2, 
\end{equation}
where $k'$ and $q=k-k'$ are the outgoing lepton and virtual boson four-momenta, respectively, 
$s=(P+k)^2$ is the centre of mass energy squared and
$W^2=s y (1-x)+m_p^2$ is the invariant mass squared of the hadronic final state, with $m_p$ the proton mass. 
The leading order NC DIS cross section for the scattering of an electron of energy $E_e$ on a proton target then reads
\begin{equation}
\label{LOeP_DIS}
\frac{d^2 \sigma^{ep \rightarrow eX}}{dx_B dQ^2}=
\frac{2 \pi \alpha_{em}^2}{Q^4} (1+(1-y)^2) \sum_q e_q^2 \, [ f_{q/p}(x_B,Q^2)+f_{\bar{q}/p}(x_B,Q^2) ]\,,
\end{equation}
where the sum runs over active quarks $q$ with electric charge $e_q$.
The differential cross section in eq.~(\ref{LOeP_DIS}) is evaluated    
by using free-nucleon, leading order, parton distributions $f_{q/p}(x_B,\mu_F^2)$ of Ref.~\cite{GRVproton}, 
setting the factorisation scale to $\mu_F^2=Q^2$. 
In order to provide a minimal quark-flavour separation, we consider both electron scattering 
on proton and deuteron targets. Cross sections on the latter are obtained by averaging cross sections
on proton and neutron targets. The latter are obtained applying isospin symmetry, \textsl{i.e.}
by exchanging $u \leftrightarrow d$ and $\bar{u} \leftrightarrow \bar{d}$ 
parton distributions in eq.~(\ref{LOeP_DIS}). 
\begin{figure}[t]
\begin{center}
\includegraphics[width=7cm,height=7cm]{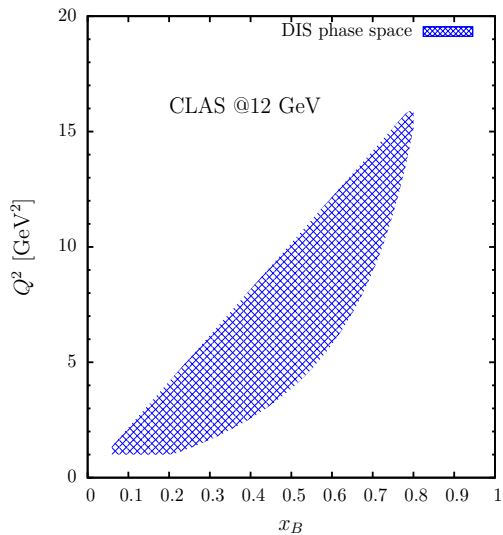}
\caption{\textit{Kinematic coverage at CLAS@12 GeV after cuts in eq.~(\ref{DIS_selection}).}}
\label{Fig:clas12_kine}
\end{center}
\end{figure}
We set the electron beam energy $E_e$ to 12 GeV.
The label $\Omega$ stands for a set of temptative cuts which define the NC DIS selection:
\begin{equation}
\label{DIS_selection}
0.2<y<0.8, \;\;\; Q^2>1 \; \mbox{GeV}^2, \;\;\; W^2>5 \; \mbox{GeV}^2\,. 
\end{equation}
The resulting phase space coverage is shown in the $(x_B,Q^2)$ plane in Fig.~(\ref{Fig:clas12_kine}).
In order to reduce the dependences on higher order corrections, all predictions 
presented in the following are normalised, if not otherwise stated, to the inclusive NC DIS cross section,
$\sigma^{\mbox{\tiny{DIS}}}_{\Omega}$, which is obtained integrating eq.~(\ref{LOeP_DIS})
over the phase space region $\Omega$ defined by constraints in eq.~(\ref{DIS_selection}).

\section{Semi-Inclusive DIS}
\label{SIDIS_Lambda}
\noindent
We consider the Semi-Inclusive process
\begin{equation}
\label{SIDISprocess}
l(k)+p(P)\rightarrow l(k')+\Lambda(h)+X\,,
\end{equation}
where, beside the scattered lepton, an additional Lambda hyperon is detected in the final state
with four-momentum $h$. Final-state hadrons produced in SIDIS are generally described  
by using the Lorentz-invariant variable 
\begin{equation}
\label{zh}
z_h=\frac{P \cdot h}{P \cdot q} = \frac{E_h^*}{E_P^*(1-x_B)} \frac{1+\cos \theta^*}{2}\,.
\end{equation}
The last equality holds in the photon-hadron centre-of-mass frame,
with the photon momentum aligned in the $+z$ direction and $\theta^*$ the hadron production angle 
with respect to the photon direction.
Hadrons produced collinearly to the spectator system have $\theta^* \simeq \pi$, so that, 
in terms of the $z_h$ variable defined in eq.~(\ref{zh}), they overlap with soft ones
(for which instead $E_h^* \simeq 0$  irrespective of the production angle) and both accumulate 
at $z_h \simeq 0$.
The $z_h$ variable defined in eq.~(\ref{zh}) is therefore well suited to described hadron production 
in the current region, but presents an ambiguity in dealing with hadrons produced by target fragmentation. 
In order to avoid this problem, cross sections can be evaluated in terms of the energy fraction 
$z_G$~\cite{graudenz} defined by
\begin{equation}
\label{zG}
z_G = \frac{E_h^*}{E_P^*(1-x_B)}= \frac{2 E_h^*}{W}, \;\; \;\;  \zeta=\frac{E_h^*}{E_P^*}\,,
\end{equation}
where $E_p^*(1-x_B)=W/2$ is the spectator energy in the photon-hadron centre-of-mass frame. Adopting such definition, 
higher order corrections can be systematically taken into account, both in the current 
and in the target fragmentation region~\cite{graudenz} so that different hadron production mechanisms are distinguished only by their peculiar $z_G$ spectrum. 
Adopting these definitions, the neutral-current semi-inclusive lowest order cross section
for producing an unpolarised Lambda off a proton in the target fragmentation region reads~\cite{graudenz}
\begin{equation}
\label{LOeP_SIDIS}
\frac{d^3 \sigma^{ep \rightarrow e\Lambda X}}{dx_B dQ^2 d\zeta}=J
\frac{2 \pi \alpha_{em}^2}{Q^4} (1+(1-y)^2)
\sum_q e_q^2 [ M_{q/p}(x_B,\zeta,Q^2)+M_{\bar{q}/p}(x_B,\zeta,Q^2) ]\,.
\end{equation}
The cross section has been expressed for later convenience 
in term of the $\zeta$ variable ($x_B+\zeta<1$) in eq.~(\ref{zG}), and the jacobian $J=\zeta[(1-x_B)|x_F|]^{-1}$ has been explicitely indicated~\cite{Mulders}.
The latter reduces to unity in the high-energy limit and it is therefore often omitted in the literature. 
In eq.~(\ref{LOeP_SIDIS}) the production of unpolarised Lambdas
in the remnant direction is described by fracture functions 
$M_{i/p}^{\Lambda}(x_B,\zeta,\mu_F^2)$~\cite{trentadue_veneziano}.
These distributions express the probability to find a parton of flavour $i$ with fractional 
momentum $x_B$ at virtuality $\mu_F^2$ in the proton conditional to the detection of a target Lambda 
with a fraction $\zeta$ of the incoming proton momentum. As for inclusive parton distributions, 
we set the factorisation scale to $\mu_F^2=Q^2$.  
In order to obtain cross sections on isoscalar target we proceed as in the inclusive DIS case,
exploiting isospin symmetry of the initial conditions. More details on this point may found 
in Sec.~\ref{Model_details}. Distributions in a given kinematic variable $v=v(x_B,Q^2,\zeta)$ are then calculated integrating the SIDIS cross section in eq.~(\ref{LOeP_SIDIS}) as follows
\begin{equation}
\label{norm_distributions}
\frac{\Delta \sigma_i^{\Lambda}}{\Delta v_i}=\frac{1}{\Delta v_i} \int_{\Omega'} dx_B \, d Q^2 \, d\zeta 
\frac{d^3 \sigma^{\Lambda}}{dx_B \, d Q^2 \, d\zeta}\, \Theta(v-v_i)\, \Theta(v_{i+1}-v)\,,
\end{equation}
where the index $i$ labels the $i$-th bin, $v_{i+1}$ and $v_{i}$ indicate
the experimental bin-edges, $\Delta v_i=v_{i+1}-v_{i}$ stands for the bin-size
and $\Omega'$ is a subset of the DIS selection, $\Omega' \subseteq \Omega$. 
\section{Model details}
\label{Model_details}
\noindent
Lambda fracture functions appearing in eq.~(\ref{LOeP_SIDIS}) have been determined 
through a global QCD fit to a variety of semi-inclusive Lambda production data in Ref.~\cite{ceccopieri_mancusi}.
In that analysis we assumed that, at an arbitrary low but still perturbative scale $Q_0^2$, 
fracture functions factorise into the product of ordinary parton distributions $f_{i/p}(x_B,Q_0^2)$ and what we address as spectator-fragmentation functions $\widetilde{D}^{\Lambda}_{i/p}(z_G)$:
\begin{equation}
\label{inputFF}
(1-x_B) \, M_{i/p}^{\Lambda}\left( x_B, \zeta ,Q^2_0 \right) =
M_{i/p}^{\Lambda}(x_B, z_G ,Q^2_0) =f_{i/p}(x_B,Q_0^2)  \widetilde{D}^{\Lambda}_{i/p}(z_G)\,,\; i=q,\bar{q},g\,.
\end{equation}
Such an assumption, supported by the fit, is motivated by considering the relevant timescales in the process. The hard scattering, controlled by parton distributions, occurs in fact on timescales $\mathcal{O}(1/Q_0)$ much shorter than the typical timescale of the fragmentation process, $\mathcal{O}(1/\Lambda_{QCD})$, controlled by spectator fragmentation functions. These initial conditions for fracture functions at $Q_0^2$ are then evolved
to scales relevant for the experiments and the parameters controlling 
$\widetilde{D}^{\Lambda}_{i/p}(z_G)$ extracted by performing a fit to available data. 
The scale $Q_0^2$ is fixed in the fit to 0.5 Ge$\mbox{V}^2$.
As discussed in Ref.~\cite{ceccopieri_mancusi}, the latter were essentially able to constrain only  
a subset of the initial conditions in eq.~(\ref{inputFF}), in particular 
the spectator fragmentation functions of valence $u$ and $d$ quarks and that of sea-quarks, for which a, flavour-independent, common function was assumed:
\begin{eqnarray}
\label{inputFF_fit}
M_{u_{v}/p}^{\Lambda}(x_B,z_G,Q_0^2)&=&f_{u_{v}/p}(x_B,Q_0^2) \widetilde{D}_{u_{v}/p}^{\Lambda}(z_G)\,, \nonumber\\
M_{d_{v}/p}^{\Lambda}(x_B,z_G,Q_0^2)&=&f_{d_{v}/p}(x_B,Q_0^2) \widetilde{D}_{d_{v}/p}^{\Lambda}(z_G)\,, \\
M_{q_{s}/p}^{\Lambda}(x_B,z_G,Q_0^2)&=&f_{q_{s}/p}(x_B,Q_0^2) \widetilde{D}_{q_{s}/p}^{\Lambda}(z_G)\,, 
\;\;\; q_{s}=u_s, d_s, \bar{u}_s, \bar{d}_s, s, \bar{s}\,. \nonumber
\end{eqnarray}
\begin{figure}[t]
\begin{center}
\includegraphics[width=7cm,height=7cm]{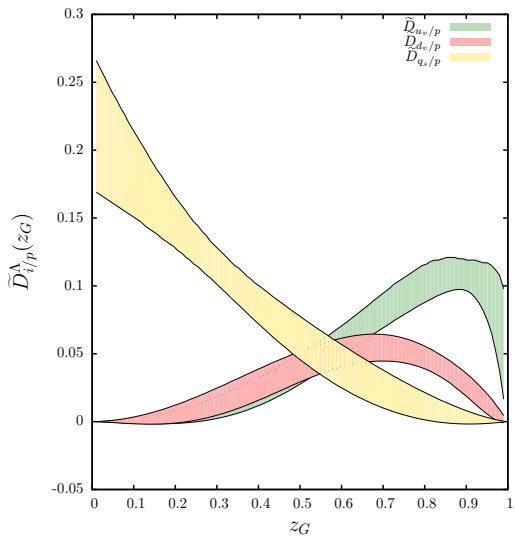}
\caption{\textit{Spectator fragmentation functions $\widetilde{D}_{i/p}^\Lambda(z_G)$ for valence $u$ (green), $d$ (red) and sea (yellow) quarks extracted from the fit of Ref.~\cite{ceccopieri_mancusi}.
The bands represent the propagation of experimental uncertainties according to the $\Delta \chi^2=1$ criterion.}}
\label{Fig:Dspect}
\end{center}
\end{figure}
Such distributions are shown in Fig.~(\ref{Fig:Dspect}) as a function of $z_G$:
the fragmenting spectrum of the $ud$-spectator into Lambdas, $\widetilde{D}^{\Lambda}_{u_v/p}$,
is harder with respect to the $uu$-one, $\widetilde{D}^{\Lambda}_{d_v/p}$. 
The leftover spectator system, $\widetilde{D}^{\Lambda}_{q_s/p}$,
associated with the hard scattering occurring on sea quarks, 
has higher Fock components and shows a softer spectrum. 
The fit was insensitive to any reasonable choice of gluon spectator fragmentation function, so
the latter was fixed to be equal to the sea-quark one,  
$\widetilde{D}_{g/p}^{\Lambda}(z_G)=\widetilde{D}_{q_{s}/p}^{\Lambda}(z_G)$.
The structure of the initial conditions allows to evaluate valence-quark fracture functions at any $Q^2$ as the difference between $M^\Lambda_{q/p}$ and $M^\Lambda_{\bar{q}/p}$. The cross sections on a neutron target 
requires the knowledge of neutron-to-Lambda fracture functions, $M^\Lambda_{i/n}$. 
We relate the latter to $M^\Lambda_{i'/p}$ assuming the following relations:
\begin{eqnarray}
\label{isospinFF}
M_{d_{v}/n}^{\Lambda}(x_B,z_G,Q^2)&=&M_{u_{v}/p}^{\Lambda}(x_B,z_G,Q^2)\,,\nonumber\\
M_{u_{v}/n}^{\Lambda}(x_B,z_G,Q^2)&=&M_{d_{v}/p}^{\Lambda}(x_B,z_G,Q^2)\,,\\
M_{q'_{s}/n}^{\Lambda}(x_B,z_G,Q^2)&=&M_{q_{s}/p}^{\Lambda}(x_B,z_G,Q^2)\,, \;\; 
q'_s=u_s,d_s,\ldots \;\; q_s=d_s,u_s,\ldots\,. \nonumber
\end{eqnarray}
The first one in eq.~(\ref{isospinFF}) appears to be natural as the quark content ($ud$) of the leftover spectator system is the same on the left and right hand side. 
The second assumption implies that the $dd$-spectator has the same fragmenting spectrum as the $uu$-one
and it does not have any physical motivation other then reducing the number of free parameters in the fit.
The third one relies on the particular flavour-symmetric choice for $\widetilde{D}_{q_s^\Lambda}(z_G)$ 
in eq.~(\ref{inputFF_fit}).  
Following the method outlined in Refs.~\cite{CTEQ,MRST} we provided, beside the best fit parametrisation,  
additional 14 Lambda fracture functions alternative parametrisations satisfying the $\Delta \chi^2=1$ criterion.
In this way experimental uncertainties can be propagated to any other observable by computing it 
for each given alternative set and then adding in quadrature the displacements with respect to best fit result.
This method has been used to obtain the error bands associated with the spectator fragmentation functions 
in Fig.~(\ref{Fig:Dspect}) and will be used in the following to estimate experimental uncertainties for the relevant cross sections. 
We close this Section mentioning that, at least in principle, we would be interested in promptly produced Lambdas.
It is well known, however, that a fraction of the measured Lambda yield comes from 
the decay of heavier resonance into Lambdas, the so-called feed-down effect.
The subtraction of these fractions from the Lambda yields was not clearly stated or even technically achievable  
in many of the experimental analyses whose data have been used in the fit of Ref.~\cite{ceccopieri_mancusi}.
We assumed therefore that the quoted yields referred to an unsubtracted Lambda sample. 
These production mechanisms are taken into account by the Lambda fracture function set of Ref.~\cite{ceccopieri_mancusi}, via effective modifications of the spectator fragmentation functions returned by the fit. 
Such an assumption must be kept in mind when comparing predictions based on the present model with forthcoming data.

\section{Predictions}
\label{pheno}
\begin{figure}[t]
\begin{center}
\includegraphics[scale=0.5]{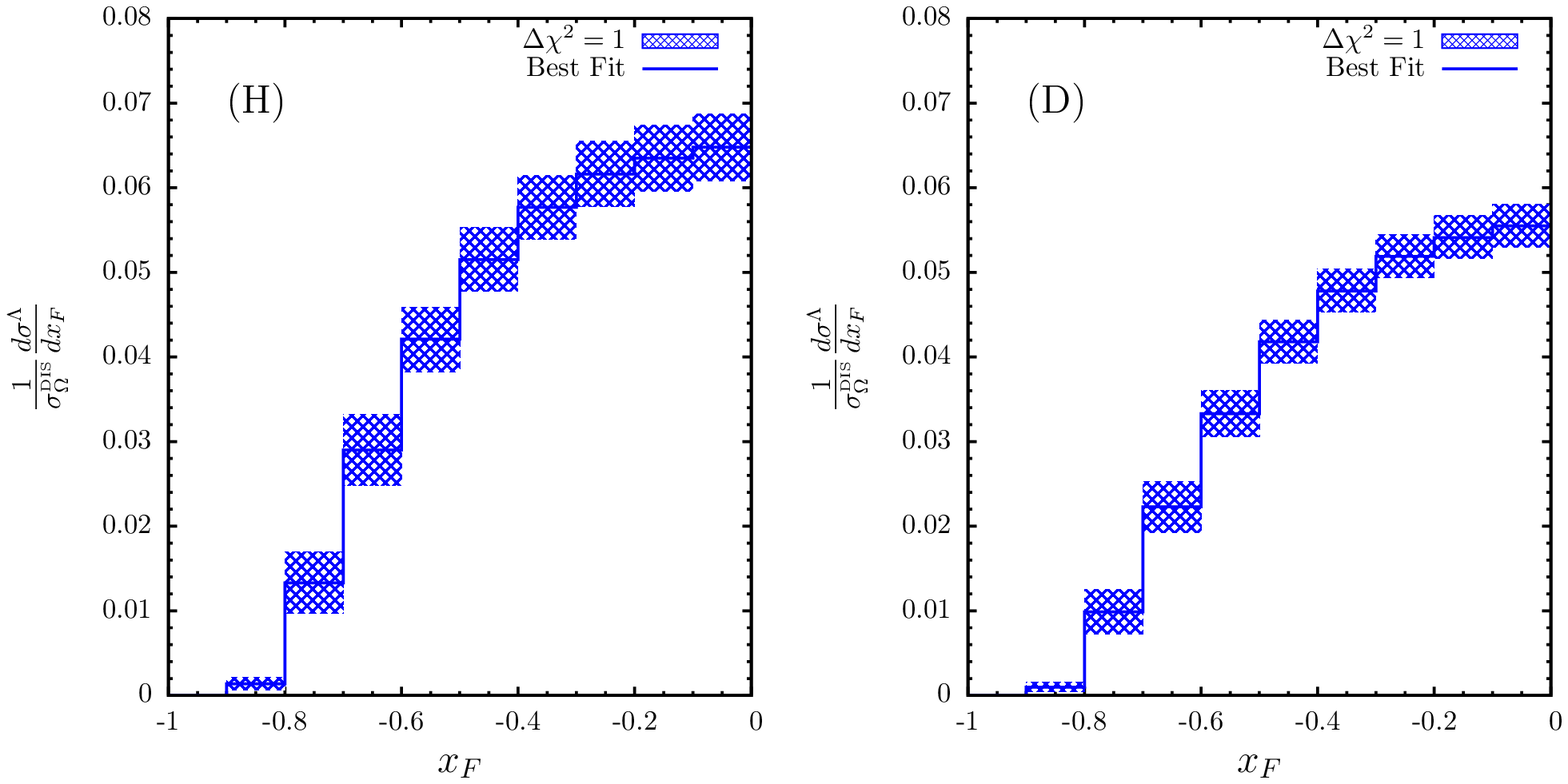}\\
\includegraphics[scale=0.5]{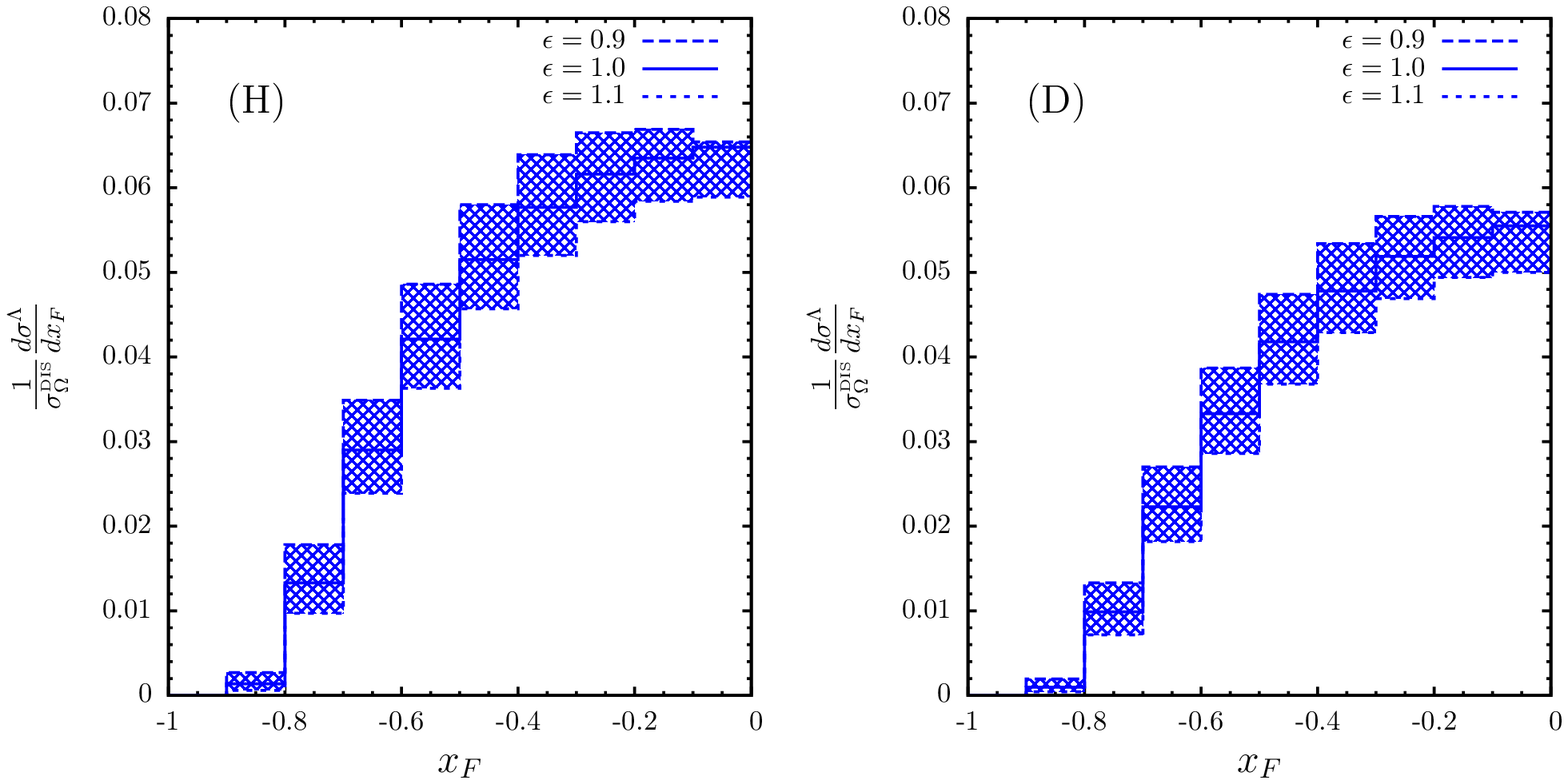}\\
\includegraphics[scale=0.5]{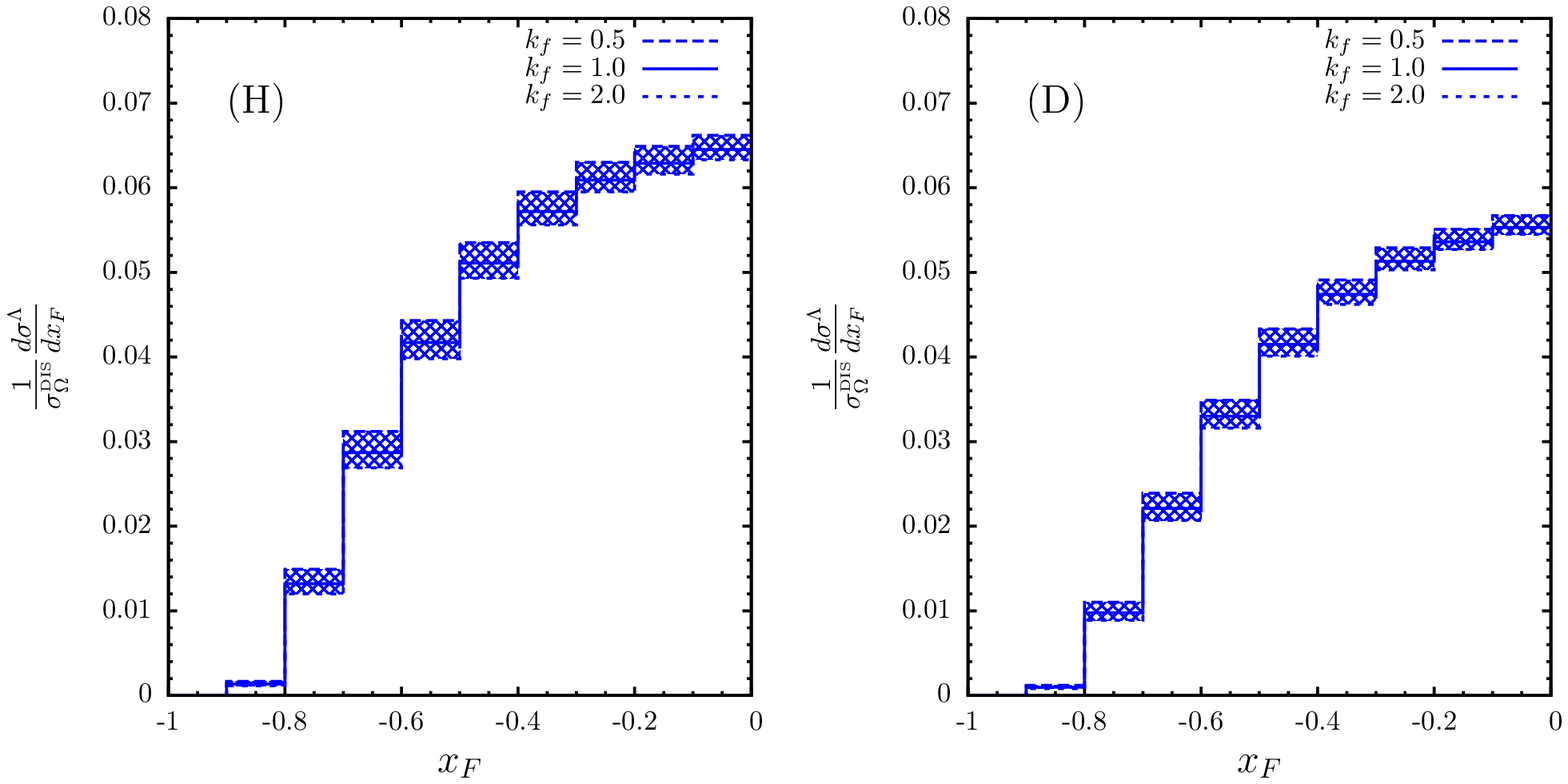}
\caption{\textit{Normalised single-differential cross sections as a function of $x_F$. Cross sections on proton (H) and deuteron (D) targets are  shown. From top to bottom the error bands represent the propagation of experimental uncertainties from the fit~\cite{ceccopieri_mancusi}, sensitivity to mass corrections and to factorisation scale variations.}} 
\label{Fig:error_cs}
\end{center}
\end{figure}

\noindent
Cross section differential in the energy ratio $z_G$
defined in eq.~(\ref{zG}) characterises the full particle production spectrum. 
The latter is given by the sum of the target fragmentation contribution, given in eq.~(\ref{LOeP_SIDIS}), and 
the current one, in which fracture functions appearing in eq.~(\ref{LOeP_SIDIS}) are replaced
by appropriate products of parton distribution and fragmentation functions. 
As discussed in Section~\ref{SIDIS_Lambda}, different hadron production mechanisms are then distinguished only by their peculiar $z_G$ spectrum and target fragmentation can be quantified without imposing any arbitrary kinematical cuts. So far, unfortunately, experimental data have not been presented in terms of this variable. 
Even in that case, however, the extraction of fracture functions with such a procedure requires an accurate knowledge of the current fragmentation contribution at low scales, whereas fragmentation functions are generally constrained at much higher scales than the ones involved in SIDIS experiments. 
In order to circumvent this problem we assumed in Ref.~\cite{ceccopieri_mancusi} that current and target fragmentation give their dominant contributions in distinct regions of space phase. Within this context it proves useful to introduce the Feynman's variable 
\begin{equation}
\label{variables3}
x_F = \pm \Big(z_G^2-\frac{4 \epsilon m_T^2}{W^2}\Big)^{\frac{1}{2}}\,,
\end{equation}
defined in the photon-proton centre-of-mass frame. The parameter $\epsilon$ will be used in the following 
to estimate the sensitivity of the predictions to Lambda mass corrections and it is fixed to $\epsilon=1$. 
We assumed that the current and target 
contributions can be kinematically separated in terms of this variable with target fragmentation 
giving its contribution for $x_F<0$ and current fragmentation for $x_F>0$. 
We stress again that the choice of $x_F=0$ as a sharp separation point is arbitrary and frame dependent. 
Moreover it is reasonable to expect that there will be an overlap 
region in which both fragmentation mechanisms will contribute, as suggested by the $\mathcal{O}(\alpha_s)$ calculation of Ref.~\cite{graudenz}.  
This strategy was adopted in Ref.~\cite{ceccopieri_mancusi} in the extraction of Lambda fracture functions
and we shall consider it as an operative choice to be tested against forthcoming data. 
We have introduced in eq.~(\ref{variables3}) the Lambda transverse mass, $m_T^2=p_{\Lambda,\perp}^2+m_{\Lambda}^2$, defined in terms of its transverse momentum and mass squared. Since it is experimentally known~\cite{NOMAD} that $\langle p_{\Lambda,\perp}^2 \rangle \ll m_\Lambda^2$, we approximate  $m_T^2 \sim m_\Lambda^2$ with $m_\Lambda=1115.683$ MeV~\cite{PDG}. Lambda-mass effects, introduced via eq.~(\ref{variables3}), are sizeable at low energies and moreover not compatible with the pQCD factorisation theorem. As described in Ref.~\cite{AKK}, such corrections are applied to the Lambda leptoproduction cross sections $\sigma^{\Lambda}$ in eq.~(\ref{LOeP_SIDIS}) via the 
extra $J$ factor. 
\begin{table}[t]
\begin{center}
\begin{tabular}{c|c} \hline \hline
 \hspace{0.4cm}  Target/Observable      \hspace{0.4cm}  & \hspace{0.8cm} $\langle n(\Lambda) \rangle$
\hspace{0.8cm} \\ \hline 
proton & 0.038   $\pm$  $0.003 (exp)^{+0.004}_{-0.004} (mass) ^{+0.002}_{-0.001} (scale) $  \\ 
deuteron & 0.032 $\pm$  $0.002 (exp)^{+0.003}_{-0.004} (mass) ^{+0.001}_{-0.001} (scale) $ \\ \hline \hline
 \hspace{0.4cm}  Target/Observable      \hspace{0.4cm}  & \hspace{0.8cm} $\sigma^{\Lambda}$ [pb]
\hspace{0.8cm} \\ \hline 
proton     & 2382    $\pm$  $170 (exp)^{+247}_{-269} (mass)^{+159}_{-125} (scale)$    \\ 
deuteron   & 1758 $\pm$  $102 (exp)^{+196}_{-206} (mass) ^{+119}_{-92} (scale)$ \\ \hline \hline
\end{tabular}
\caption{\textit{Predicted Lambda yields and cross section for $x_F<0$ on proton and 
deuteron targets. Quoted errors represent the propagation of experimental uncertanties from the fit~\cite{ceccopieri_mancusi} (exp), the expexted sensitivities to mass corrections (mass) and to factorisation scale variations  (scale).}}
\label{cs}
\end{center}
\end{table}
We begin our overview of results presenting in Tab.~(\ref{cs}) predicted yields
and absolute cross sections for the production of Lambdas with $x_F<0$ within the DIS selection
defined in eq.~(\ref{DIS_selection}). Results on proton and deuteron targets are shown. The 
former are larger than the latter, since, as already mentioned, our model 
returns  $\widetilde{D}^{\Lambda}_{u_v/p}$ larger and harder than $\widetilde{D}^{\Lambda}_{d_v/p}$.
Both the yields and the absolute cross sections are supplemented by errors. The first one 
corresponds to the propagation of experimental uncertanties coming from the fit~\cite{ceccopieri_mancusi}
and it is denoted with the label (exp). It amounts to an average uncertainty of 7\% on the yields.
Among theoretical errors we address the sensitivity to Lambda mass corrections and higher order corrections.
The former is assessed by varying, arbitrarily, the parameter $\epsilon$ appearing in eq.~(\ref{variables3}) 
in the range $\epsilon \in [0.9,1.1]$ and it is indicated with the label (mass) in Tab.~(\ref{cs}).
The latter is assessed, as costumary, by varying the factorisation scale $\mu_F^2$ both 
in fracture and ordinary parton distributions in the range $\mu_F^2 \in [0.5 \, Q^2, 2 \, Q^2]$ 
and it is indicated with the label (scale) in Tab.~(\ref{cs}).
From these numbers it appears that there is, given the relatively low beam energy of the experiment, 
a rather large sensitivity to Lambda mass corrections. 
On the other hand, errors associated to estimated higher order corrections are smaller than 
experimental uncertainties. The yields appears to be particularly stable against scale variations since 
the factorisation scale is simultaneously varied both in the numerator (\textsl{i.e.} fracture functions) and denominator ( \textsl{i.e.} inclusive parton distributions). We stress here that while the quoted experimental uncertainties have a precise statistical meaning the other two must be considered as temptative estimations of systematics errors associated with theoretical predicitions. We present in Fig.~(\ref{Fig:error_cs}) the normalised $x_F$ spectrum supplemented by the correponding uncertanties.  
\begin{figure}[t]
\begin{center}
\includegraphics[scale=0.9]{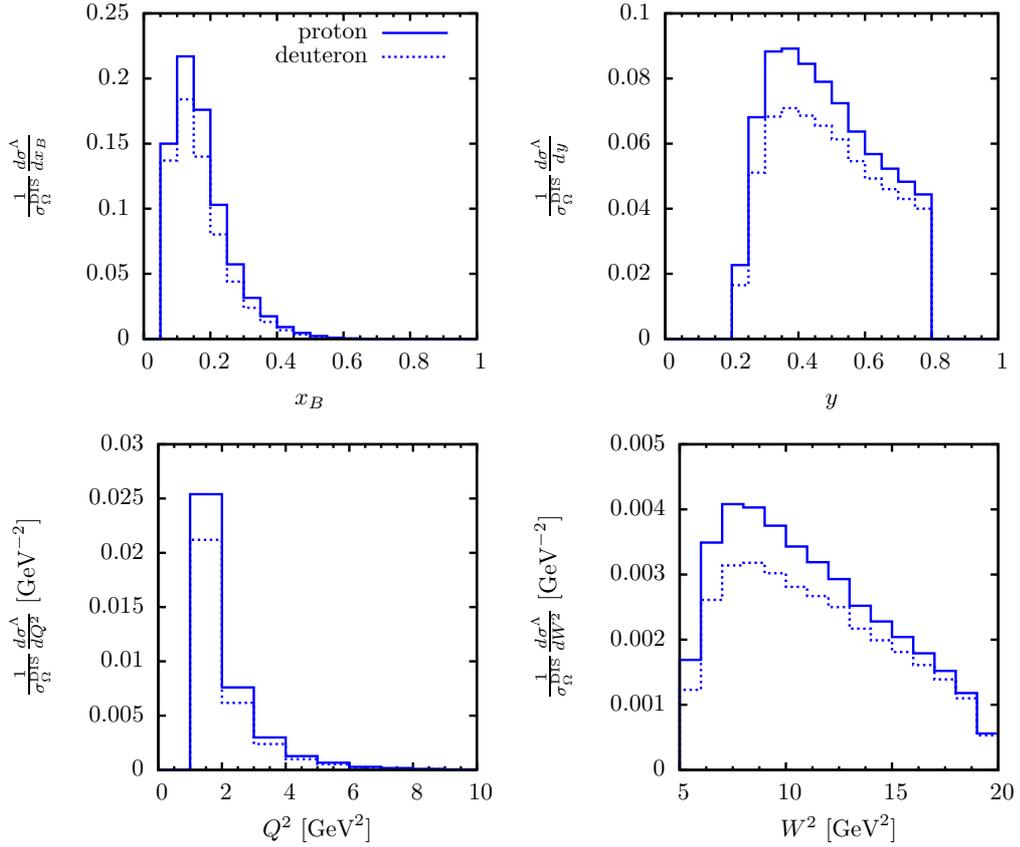}
\caption{\textit{Normalised single-differential cross sections as a function of leptonic kinematic variables
with the additional requirement of detecting a Lambda with $x_F<0$. Cross sections on proton and deuteron targets are  shown.}} 
\label{Fig:single_diff_cs}
\end{center}
\end{figure}
We now turn our attention to the production properties of backward Lambdas. 
The averaged values of kinematical variables for Lambda production in NC DIS with $x_F<0$ 
are given by $\langle x_B \rangle=0.17 \; (0.18)$,   $\langle y \rangle=0.48 \;(0.45)$,
$\langle Q^2\rangle=2.0 \;(1.9)$ Ge$\mbox{V}^2$ and $\langle W^2\rangle=11.1 \; (10.3)$ Ge$\mbox{V}^2$.
In parenthesis we have indicated the corresponding values for the inclusive DIS case. 
The values for the two processes are quite close to each other, a feature which should be ascribed 
to the factorised ansatz for fracture functions at the lowest scale in eq.~(\ref{inputFF})
which is almost preserved by evolution from $Q_0^2$ to $\langle Q^2\rangle$.  
In Fig.~(\ref{Fig:single_diff_cs}) we present the normalised single-differential cross sections as a function of scattered lepton variables with the additional requirement of detecting a Lambda in target region, $x_F<0$. Such distributions show a qualitatively similar shape irrespective of the target considered, either proton or deuteron. The $x_B$ distributions peaks at the lowest accessible values of $x_B$, given by the boundary in Fig.~(\ref{Fig:clas12_kine}). The bulk of the cross section resedes, as expected, at very low values of $Q^2$ and $W^2$ invariants.
The combined study of such distributions together with the corresponding ones in inclusive DIS can potentially highlight correlations between the hard scattering and the spectator fragmentation into target Lambdas.
\begin{figure}[t]
\begin{center}
\includegraphics[scale=0.7]{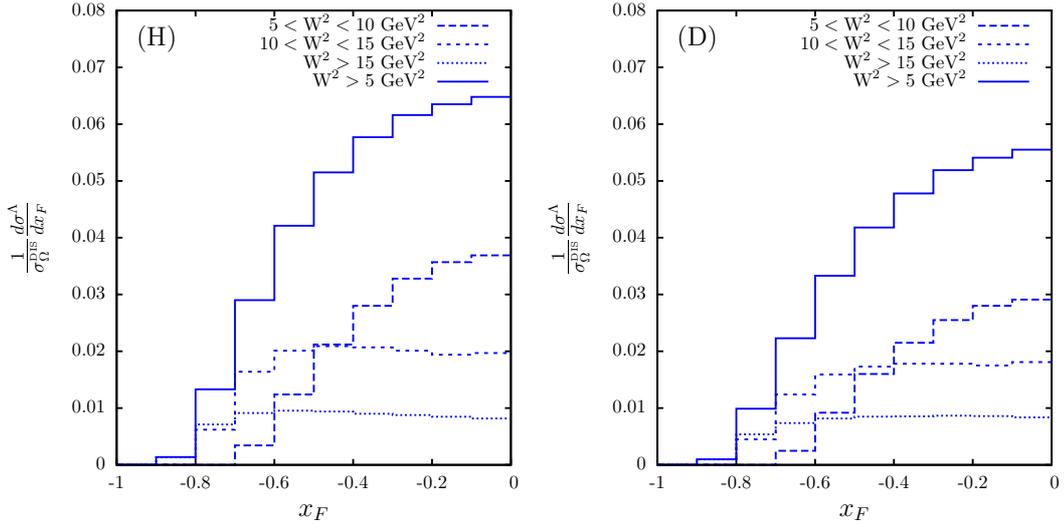}
\caption{\textit{Normalised Lambda single-differential cross section as a function of $x_F$ integrated in different range of $W^2$ on proton (H) and deuteron (D) targets.}} 
\label{Fig:W2}
\end{center}
\end{figure}
In Fig.~(\ref{Fig:W2}) we present the Lambda single-differential cross section as a function of $x_F$, 
integrated in ranges of $W^2$. The very backward production regime (at large and negative $x_F$) is accessed only at highest values of $W^2$. This is the combined effect of hadron mass corrections, via eq.~(\ref{variables3}), and the energy spectrum of the spectator fragmentation functions $\widetilde{D}_{i/p}^{\Lambda}$ shown in Fig.~(\ref{Fig:Dspect}).
\begin{figure}[t]
\begin{center}
\includegraphics[scale=0.7]{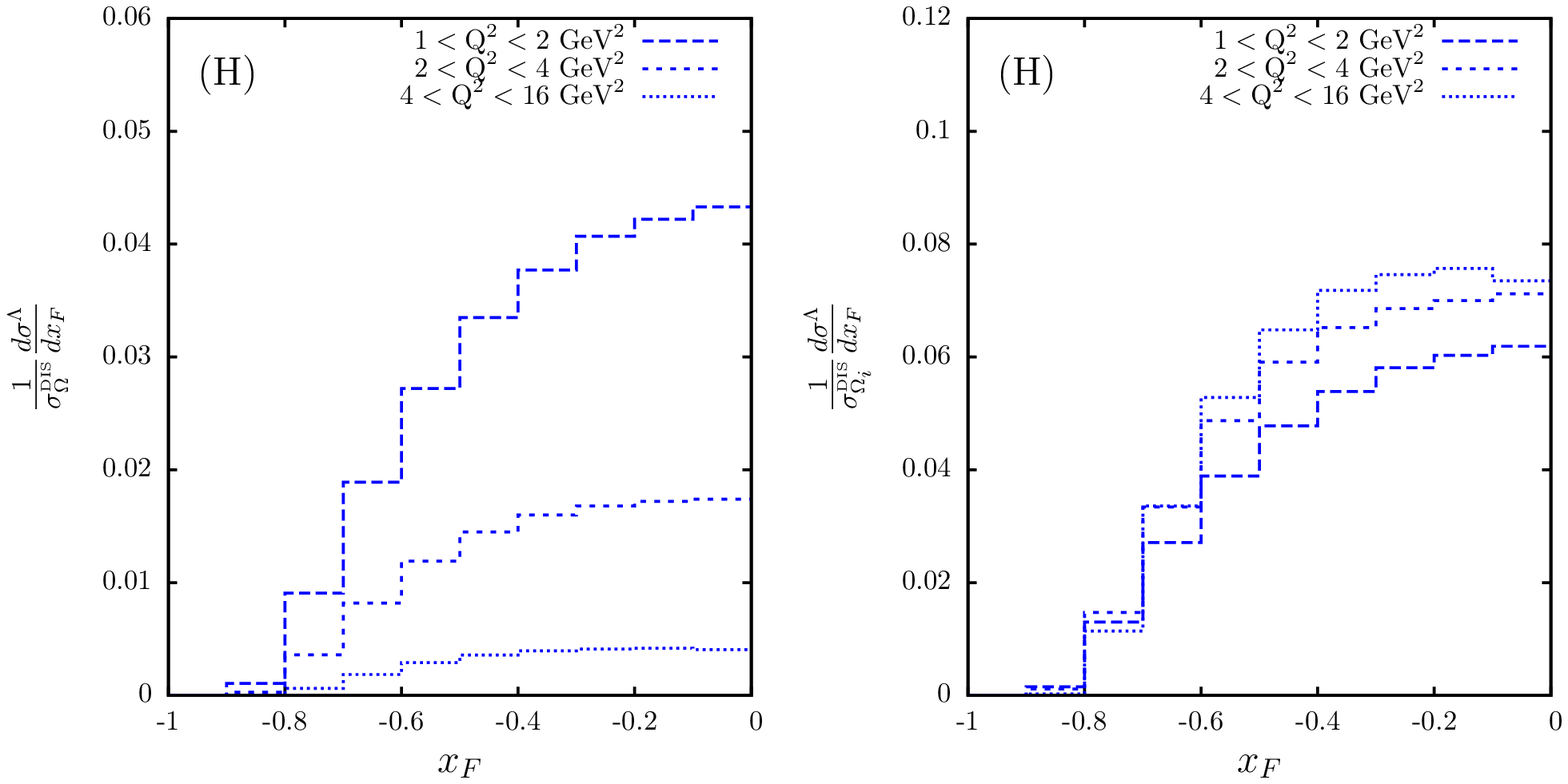}\\
\includegraphics[scale=0.7]{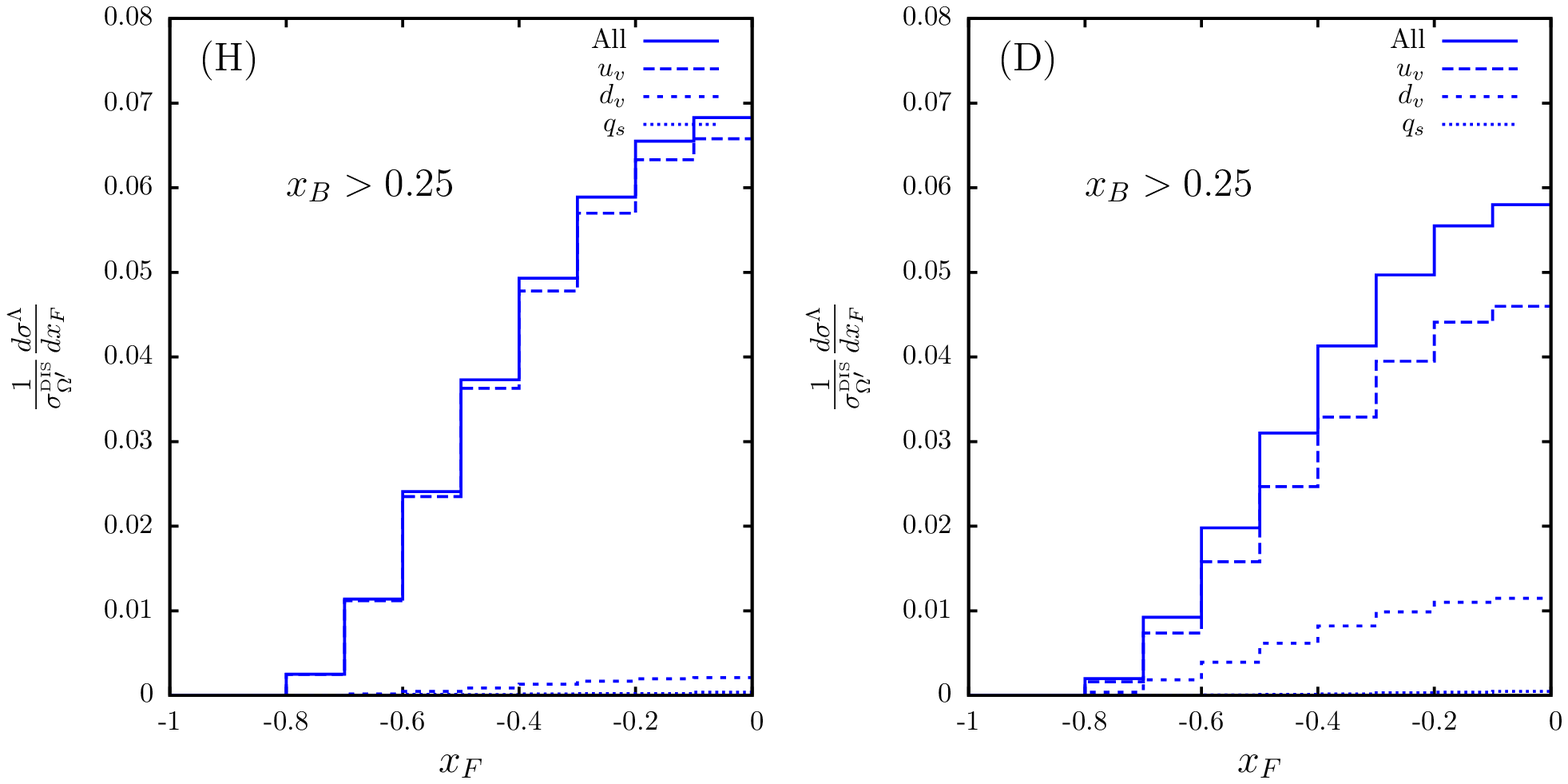}
\caption{\textit{Top left: Normalised Lambda single differential cross section as a function of $x_F$ in different range of $Q^2$ on a proton target. Top right: Lambda multiplicities as a function of $x_F$ in ranges of $Q^2$. 
Bottom: quark-flavour decomposition of the Lambda single-differential cross section as a function of $x_F$
on a proton (left) and deuteron (right) target with the additional cut $x_B>0.25$ imposed.}} 
\label{Fig:Q2_and_flavour}
\end{center}
\end{figure}
In the top left panel of Fig.~(\ref{Fig:Q2_and_flavour}) we show the $x_F$ dependence of the cross section 
in ranges of $Q^2$. As already seen in Fig.~(\ref{Fig:single_diff_cs}), the bulk of the cross section 
is at low $Q^2$, althought it remains non-negiglible to the highest accessible $Q^2$. 
\begin{figure}[t]
\begin{center}
\includegraphics[scale=0.7]{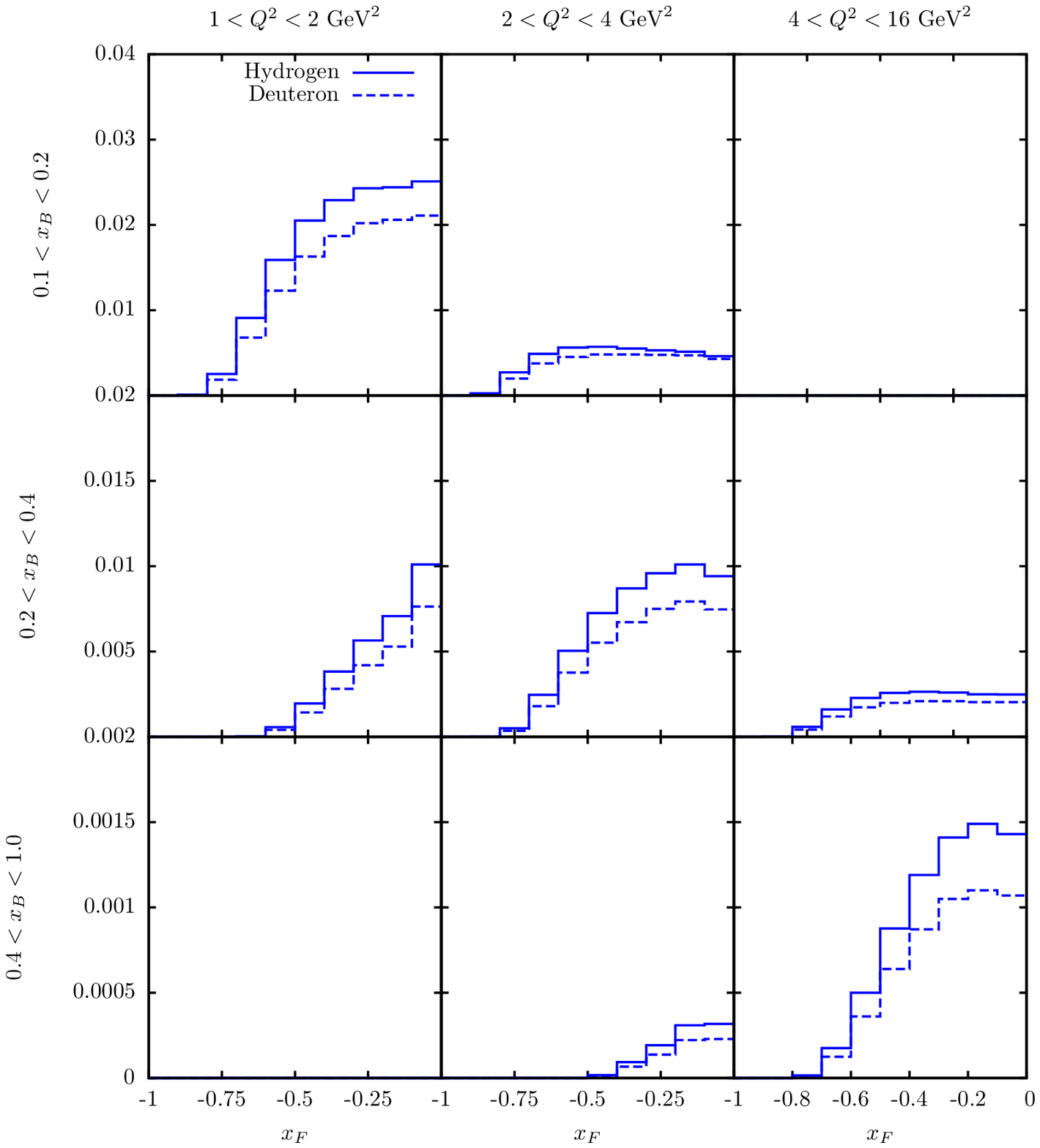}
\caption{\textit{Normalised single differential cross sections as a function of $x_F$,
$1/\sigma^{\mbox{\tiny{DIS}}}_{\Omega} d\sigma^\Lambda / dx_F$, in various bins of $x_B$ and $Q^2$. Distributions for hydrogen and deuteron targets are presented and correspondingly normalised.}} 
\label{Fig:dsdxf_bins_xbQ2}
\end{center}
\end{figure}
The measurement of the Semi-Inclusive cross section in the target region as a function of $Q^2$
can validate the leading twist nature of particle production in this region of phase space,
as assumed by fracture functions formalism. In the top right panel of the same figure we present 
Lambda multiplicities in ranges of $Q^2$ as a function of $x_F$.  In this case distributions are normalised to $\sigma^{\mbox{\tiny{DIS}}}_{\Omega_i}$, where the additional index $i=1,2,3$ stands for the corresponding $Q^2$ range
indicated on the plot which supplements the DIS selection $\Omega$.    
A mild rise of the multiplicity can be observed as $Q^2$ increases, which can be possibly ascribed to 
the QCD evolution of fracture functions. It would be extremely interesting 
to compare these distributions with the corresponding one in photoproduction regime
in order to determine to which extent the transition to the non-perturbative regime in $Q^2$ affects
the Lambda spectrum in the target region.  
In the bottom row of Fig.~(\ref{Fig:Q2_and_flavour}) we show the quark-flavour decomposition 
of the single differential cross sections as a function of $x_F$ normalised to $\sigma^{\mbox{\tiny{DIS}}}_{\Omega'}$,
where ${\Omega'}$ stands for the DIS selection $\Omega$ supplemented with the cut $x_B>0.25$.
Assuming that the experiment can be performed both on proton and deuteron targets, 
the plots show that such a selection can provide an optimal 
valence quark-flavour discrimination for Lambda fracture functions.
In the proton target case the Lambda spectrum is dominated by scattering on valence $u$-quarks
with maximal sensitivity to the fragmentation of the $ud$-spectator system into Lambdas.
In the deuteron case, the generalised isospin relations in eqs.~(\ref{isospinFF}) 
allow the extraction of the $uu$-spectator fragmentation functions. 
\begin{figure}
\begin{center}
\includegraphics[scale=0.7]{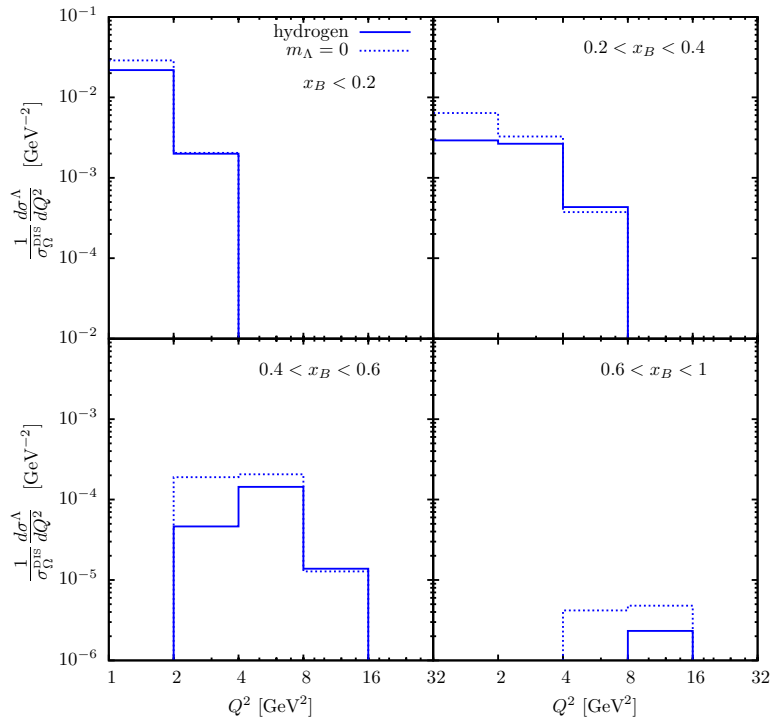}
\caption{\textit{Normalised Lambda single-differential cross section on a proton target as a function of $Q^2$ in different bins of $x_B$ and integrated in the range $x_F<0$.}}
\label{Fig:6b}
\end{center}
\end{figure}
In Fig.~(\ref{Fig:dsdxf_bins_xbQ2}) the normalised single differential cross sections as a function of $x_F$
are presented in $x_B$ and $Q^2$ bins. This way of presenting the data
is probably the more exhaustive and it might be valuable for the determination of Lambda fracture functions in forthcoming global fit analyses.
We conclude this Section presenting in Fig.~(\ref{Fig:6b}) the normalised Lambda single-differential cross sections on a proton target as a function of $Q^2$ in four different bins of $x_B$. The final-state Lambda is required to have $x_F<0$. The $Q^2$-differential cross section deserves special attention 
since  this observable may provide crucial test for the predicted evolution of fracture functions and validate the key assumptions of the underlying theory. Given the relatively low values of $W^2$ accessed by the experiment, 
the $Q^2$ spectrum shows significant hadron mass corrections, as can be inferred comparing 
default predictions with the one in which the Lambda mass has been set to zero. 
Their effect is to suppress the cross section as $x_B$ increases. 
In view of these results, the genuine $Q^2$ dependence of the cross section and mild logarithmic 
effects generated by QCD evolution of fracture functions 
can get obscured by hadron mass corrections. 
Therefore the interpretation of forthcoming data will require 
a proper modelisation of the latter either with the basic method described 
in this paper or with more refined treatment as the one discussed in Ref.~\cite{Accardi}.

\section{Conclusions} 
\label{Conclusions}
In this paper we have considered Lambda production in the target fragmentation region of electron-proton deep inelastic scattering. We have presented, based on a recently obtained set of Lambda fracture functions, predictions for a number 
of relevant observables supplemented with a conservative error estimates.
In future perspective, the subdivision of $x_F$ spectra in bins of $Q^2$ and $x_B$ can be a valuable input for forthcoming fits. Given the energy range of the considered experiment, the possibility to use different light targets offer an additional handle on Lambda fracture functions quark-flavour separation in the valence region.
The study of the of the $Q^2$ dependence of the cross sections can be valuable to test and validate 
the key feature of the underlying theory and many of the assumptions of the proposed phenomenological model.

\end{document}